# Axonal Conduction Velocity Impacts Neuronal Network Oscillations

Vladimir A. Ivanov
The Computational Brain Lab
Rutgers, The State University of New Jersey
New Brunswick, NJ 08854 USA
vladimir.ivanov@rutgers.edu

Ioannis E. Polykretis
The Computational Brain Lab
Rutgers, The State University of New Jersey
New Brunswick, NJ 08854 USA
ip211@cs.rutgers.edu

Konstantinos P. Michmizos
The Computational Brain Lab
Rutgers, The State University of New Jersey
New Brunswick, NJ 08854 USA
konstantinos.michmizos@cs.rutgers.edu

*Abstract*— Increasing experimental evidence suggests that axonal action potential conduction velocity is a highly adaptive parameter in the adult central nervous system. Yet, the effects of this newfound plasticity on global brain dynamics is poorly understood. In this work, we analyzed oscillations in biologically plausible neuronal networks with different conduction velocity distributions. Changes of $1-2$ (ms) in network mean signal transmission time resulted in substantial network oscillation frequency changes ranging in $0-120$ (Hz). Our results suggest that changes in axonal conduction velocity may significantly affect both the frequency and synchrony of brain rhythms, which have well established connections to learning, memory, and other cognitive processes.

## I. Introduction

The timing of action potential (AP) arrival is of great importance to information processing in brain circuits [1]. Experimental studies have revealed that a number of pathways, such as thalamic pathways to the cortex or the auditory brainstem, maintain fine-tuned spike time arrival, with sub-millisecond precision [2,3]. Individual axons primarily control spike time arrival through their AP conduction velocity, which is modifiable through various mechanisms including ion channel densities, axonal structure, and myelinating glia, which wrap their processes around axons and form the myelin sheath that changes AP propagation speeds [1]. The prevailing hypothesis has always been that these mechanisms actively shape neuronal circuits during development of the central nervous system (CNS) [4]. Aside from slow homeostatic adjustments, conduction velocities in the adult brain have been considered static [4]. However, the emergence of new visualization tools has revealed that AP conduction velocities are highly adaptive in adult neuronal circuits [4].

Axons exhibit conduction velocity plasticity in response to neuronal activity through multiple mechanisms operating at different time scales [1]. Unmyelinated axons can adjust their conduction velocity through axon depolarization and diameter adjustment, while myelinated axons tend to rely more on their interaction with the myelin sheath [5,6]. Interestingly, some of these mechanisms rapidly alter conduction velocity in a matter of seconds or minutes. For example, depolarization of an unmyelinated axonal cell membrane, by prior APs, results in conduction velocity slowing within a matter of minutes [5]. Similarly, depolarization of oligodendrocytes, the primary myelinating glial cell in the CNS, increases conduction velocity in the axons they myelinate from 20 seconds to 3 hours [9]. Moreover, new myelin formation is driven by local neuronal activity, thereby changing local axonal conduction velocities over the course of several hours [10]. This last form of myelin plasticity has been shown to play a critical role in motor learning, suggesting that the effects of adaptive axonal AP conduction velocities may translate to network level dynamics.

In this work, we computationally investigated the effects of axonal AP conduction velocity distributions on the oscillatory behavior of neuronal networks. To do so we developed an Izhikevich-type network that conformed to several generally observed characteristics of cortical networks including the ratio between excitatory and inhibitory neurons, sparseness of connectivity, local interneuron inhibition, and lognormal synaptic weight distributions [8,11,13]. Most essential to our goal, is that such an Izhikevich network has been shown to exhibit brain-like rhythms [12]. Our analysis revealed multiple, nontrivial relationships between network conduction velocity statistics and the corresponding network oscillation frequency, and synchrony level. Our results suggest a fascinating possibility that neuronal networks may change their oscillatory activity in response to precisely tuned AP propagation delays.

## II. Methods

We developed Izhikevich-type neuronal networks with all parameters being static. Each network was initialized and simulated with a different conduction velocity distribution that was biologically constrained within the reported range [14]. The networks were analyzed for the ability of all their neurons to get entrained into synchronized activity, namely oscillation frequencies and levels(extent) of neuronal synchrony.

Our network architecture followed the one previously reported in [12]. Briefly, our network consisted of 1000 Izhikevich neurons, a simple, semi-empirical model of a cortical neuron [12]. The neuronal membrane fluctuation was described by a set of two differential equations:

Ioannis E. Polykretis was partially funded by the Onassis Foundation Scholarship

$$dv = 0.04v^2 + 5v + 140 - u + I \qquad (1)$$
$$du = a(bv - u)$$

where $v, u \in \mathcal{R}$ are the fast activation and slow recovery variables, and $a, b, c, d \in \mathcal{R}$ determined neuron type. Neurons spiked if $v > 30$ (mV), and variables $v, u$ were reset as,

$$v = c \qquad (2)$$
$$u = u + d$$

The neuronal population consisted of 80% excitatory, regular spiking (RS), and 20% inhibitory, fast spiking (FS), neuron types, per experimental data [11]. Neuron parameters were obtained from [12]. Network connectivity depended on neuron type, with each neuron uniformly connected to 10% of the network [8]. Outgoing connections from excitatory neurons were connected to both excitatory and inhibitory neurons, while inhibitory neurons were connected only to excitatory neurons. We initialized excitatory and inhibitory synaptic weights with $(\mu, \sigma^2)_{exc} = (1.67, 0.5)$ and $(\mu, \sigma^2)_{inh} = (1.49, 0.5)$ lognormal distributions, preserving the experimentally observed functional form of the distribution [13]. Lastly, axonal AP conduction velocities were modeled as discrete time delays, with a resolution of 1 (ms); same as the network integration timestep. All inhibitory connections were assigned the minimum delay of 1 (ms), mimicking local interneuron inhibition [11]. Experimental data shows that the distributions of cortico-cortical axonal AP conduction delays from various cortical areas to the midline can be approximated using normal distributions described by moments ranging in $\mu = [3,6]$ (ms) and $\sigma^2 = [0.5, 5.0]$ with all measured delays largely limited to $0 - 10$ (ms) [14]. We doubled this range to $0 - 20$ (ms) to account for the fact that experimental data measured pathway delays only up to the midline, or half of possible total length. Therefore, for simulation purposes we restricted excitatory conduction delay distributions to normal distributions with $\mu \in [0, 20]$ (ms) and $\sigma^2 \in [0.5, 5.0]$.

The purpose of the simulation process was to relate conduction delay distribution to global network dynamics. Simulations consisted of four steps, repeated for each unique delay distribution. First, a neuronal network was generated with a unique conduction delay distribution using the previously described rules and statistics. Second, the network was launched and stimulated continuously for 5 (s). Stimulation consisted of randomly selecting a neuron every millisecond and injecting 20 (pA) of current into it. During network run time, all network parameters remained constant. Third, spike data was collected for the last 4 (s) of simulation time. This simulation process was repeated several times for each conduction delay distribution simulated to ensure robustness and generalization of our results. We found the results were repeatable with little to no variation. Lastly, spectral and synchronization analysis were performed on the recorded data.

Network oscillation frequency was obtained from recorded network spike data through spectral analysis. First, each simulated network's 2-dimensional, $800 \times 4000$, spike data was averaged over all neurons resulting in the 1-dimensional network activity time series, $S(t)$, signal. We then applied the Fast Fourier Transform (FFT) algorithm to $1000$ (ms) segments of the $S(t)$ signal, to extract its frequency components. Lastly, the largest frequency component was taken to represent the oscillation frequency of the corresponding network.

Network synchrony was measured on spike data using a metric based on variance of time-averaged, neuronal spike fluctuations [15]. This measure computes the variance of network level fluctuations, normalized by the average variance of fluctuations of individual neurons, as described by the following equation:

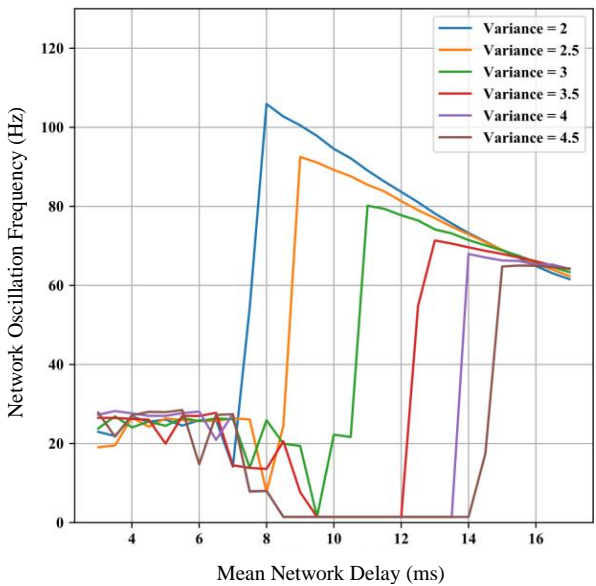

**Fig. 1.** Network oscillation frequency as a function of connection delay average, displayed for delay variances, $\sigma^2 \geq 2$. The relationship is nonlinear and step-like, with delay variance controlling its shift.

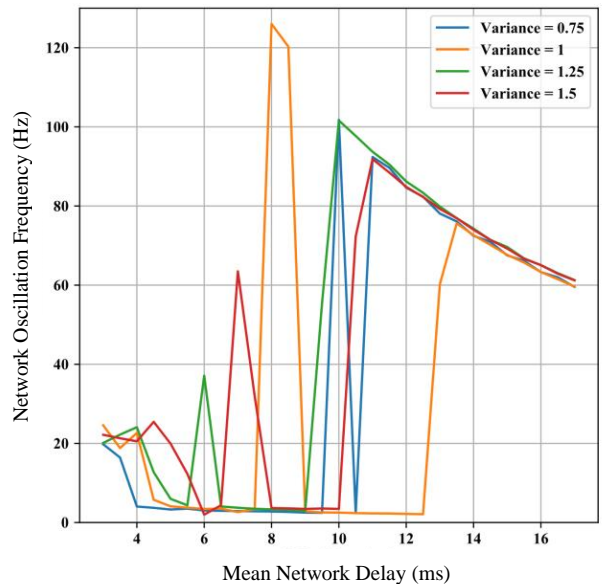

**Fig. 2.** Network oscillation frequency as a function of connection delay average, displayed for delay variances, $\sigma^2 < 2$. The relationship is highly nonlinear with a double peak form. Delay variance both shifted and noticeably altered the response curve.

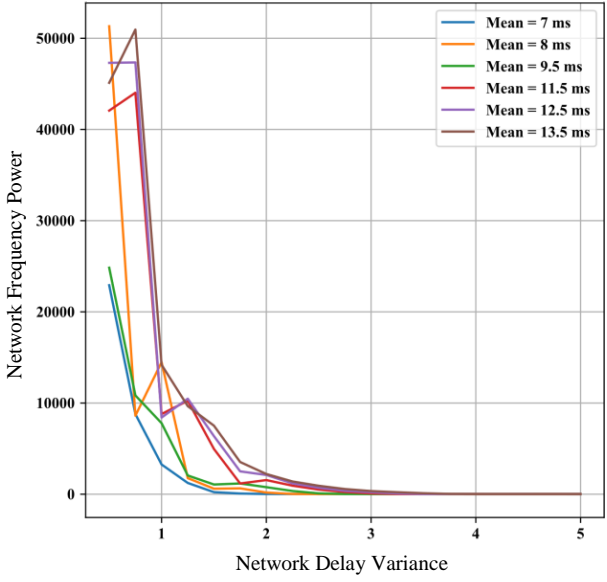

**Fig. 3.** Relative power of network oscillation frequency as a function of delay variance. Lower variance values resulted in higher relative power levels and vice versa.

$$\chi_n = \frac{\sigma_S^2}{\sigma_{S_i^2}} = \frac{\langle [S(t)]^2 \rangle_t - [\langle S(t) \rangle_t]^2}{\frac{1}{N}\sum \langle [S_i(t)]^2 \rangle_t - [\langle S_i(t) \rangle_t]^2} \quad (3)$$

where $\langle \ldots \rangle_t$ represents an average over time, $S_i(t)$ is the spike train over time $t$ for neuron $i$, and $S(t)$ is the network spike train signal averaged over $N$ neurons. This measure positively quantifies network synchrony on a scale 0 to 1.

### III. RESULTS

We analyzed the effects of different axonal conduction delay distributions on network oscillations and synchronization. Our results revealed a nonlinear relationship between network oscillation frequency and mean delay of network connections. Additionally, network delay variance impacted both the form of the oscillation frequency response and network synchronization.

Network oscillation frequency exhibited a highly sensitive response to the mean delay of network connections. This nonlinear relationship is depicted in Fig. 1 and 2, depicting network oscillation as a function of average network delay. Oscillations ranged in the biologically relevant 0 - 120 (Hz). Interestingly, sub-millisecond changes in mean delay produced frequency changes on the order of tens of Hertz, shown in Fig. 1 and 2. For example, more than 80 (Hz) of oscillation frequency range is covered by a mere $1 - 2$ (ms) change in network mean conduction delay.

Variance of network delay distribution appeared to control both the position and form of the network oscillation frequency response curve. First, variance determined the shift of the response curve along the mean delay axis. Fig. 1 demonstrates this relationship most clearly, where higher variance values tend to shift the response curve towards higher mean delay values and lower frequency ranges. Conversely, Fig. 2 shows that for variance values, $\sigma^2 < 2$, the form of the response curve completely changes resulting in multiple high frequency peaks. Interestingly, the low variance regime resulted in highest network frequency of approximately 125 (Hz). Additionally, variance impacted network synchrony.

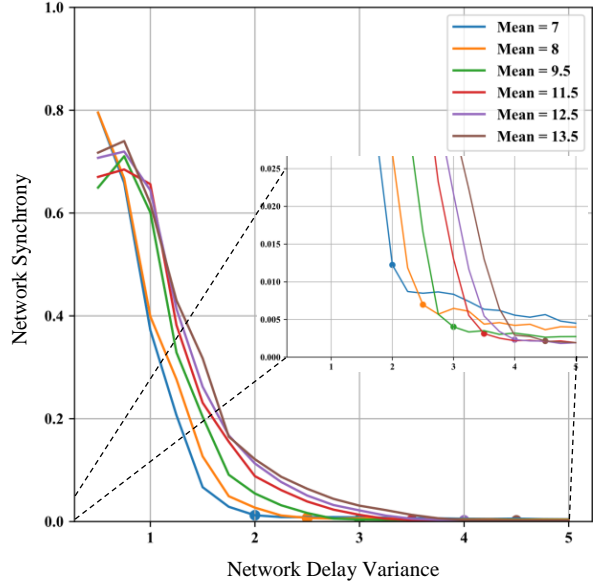

**Fig. 4.** Extent of network synchrony as a function of delay variance. Lower variance values resulted in higher increased synchrony and vice versa. (Inset) Each point corresponds to the rapid network frequency change shown in (Fig. 1).

Our model exhibited an inverse relationship between the variance of network delay distributions and the degree of network synchronization. This phenomenon was initially observed through network oscillation frequency power levels and network raster plots. Fig. 3 shows that the relative power level increases with decreasing delay variance, suggesting that network dynamics tended towards greater order with minor frequency components becoming less pronounced. Analysis of individual network raster plots confirmed this through large amplitude oscillations for low delay variance values, and small amplitudes for higher delay variance values, shown in Fig. 5. To conclusively verify this relationship, we measured network synchrony, shown in Fig. 4. Network synchrony appeared to be inversely related to delay variance, while the mean delay positively shifted the synchrony curve. Interestingly, for delay distributions with variance, $\sigma^2 \geq 2$, the abrupt rise in network frequency occurred precisely at the cusp of rapidly increasing synchronization. This can be seen in the inset of Fig. 4, where the point on each curve corresponds to the mean at which network frequency rapidly rose as seen in Fig. 1.

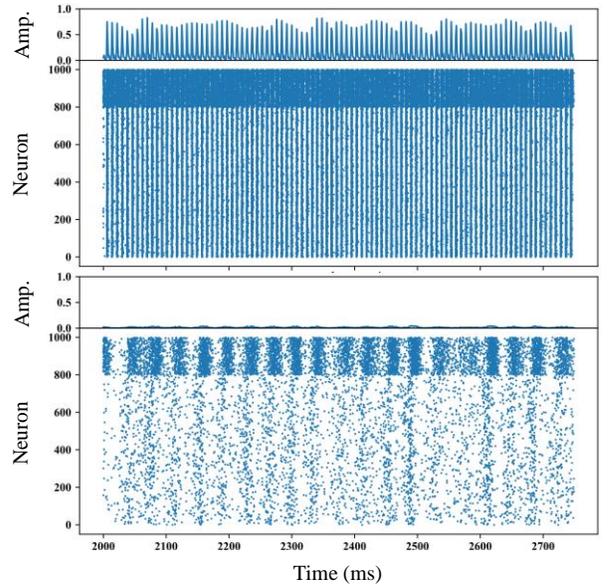

**Fig. 5.** Raster plots comparing network oscillations and amplitude for two different AP conduction delay distributions. [Top] Delay distribution of $(\mu, \sigma^2) = (8,1)$, resulted in network oscillation frequency $\approx 120$ (Hz) with high network activity. [Bottom] Delay distribution of $(\mu, \sigma^2) = (8,3)$, resulted in network oscillation frequency $\approx 35$ (Hz) with low network activity.

IV. DISCUSSION

In this paper, we demonstrated the existence of a nonlinear relationship between neuronal network conduction delay distribution and network oscillation frequency, and synchronization. Both the distribution mean and variance inflicted non-trivial effects on network behavior. Conforming to experimental evidence our model proposed a new perspective on the origins of brain rhythms.

Our modeling results suggested that the entire biologically observed network oscillation frequency range of approximately $0 - 100$ (Hz), could be partly driven by precise, sub-millisecond changes in the neuronal network's average axonal conduction delay. Intriguingly, experimental evidence corroborates such precise tuning in neuronal signal transmission, where it has been shown that the AP transmission speed between any two specific neurons is maintained at the sub-millisecond time scale with high degree of reproducibility [1]. This hints at the possibility that adaptive signal velocity mechanisms play a significant role in observed network level phenomena.

Our computational results further suggested a fascinating possibility that AP propagation speeds impact global network dynamics. Given that oscillations and synchronization are fundamental components of information processing in the brain [16], understanding the role that neuronal and non-neuronal cells have in higher cognitive functions is crucial. This challenges the long-held notion of glial passivity in information processing and reveals potential roles for non-neuronal cells proposed by us and others [18,19]. For instance, since oligodendrocytes are now known to adaptively affect AP velocity, through actively restructuring white matter [1,17], this study paves the way for computationally studying the interaction of neuronal and non-neuronal cells in brain health and disease.

This work supports our ongoing effort to investigate network AP conduction velocity distributions in the context of other network parameters such as connectivity, excitation vs. inhibition ratio, and synaptic weight distributions all of which are known to affect network level properties. This will enable us to study the prevalence of our findings in more comprehensive models of brain cells and networks.

REFERENCES

[1] Seidl, A. H., & Bloedel, V. M. (2014). Regulation of Conduction Time along Axons. *Neuroscience*, *0*, 126–134.

[2] Salami M., Itami C., Tsumoto T., & Kimura F. (2003). Change of conduction velocity by regional myelination yields constant latency irrespective of distance between thalamus and cortex. *Proceedings of the National Academy of Sciences of the United States of America*, *100*, 6174–6179.

[3] Seidl, A. H., Rubel, E. W., & Harris, D. M. (2010). Mechanisms for adjusting interaural time differences to achieve binaural coincidence detection. *The Journal of Neuroscience : The Official Journal of the Society for Neuroscience*, *30*(1), 70–80.

[4] Debanne D., Campanac E., Bialowas A., Carlier E., & Alcaraz G .(2011) Axon physiology. *Physiol Rev* 91(2):555–602.

[5] De Col, R., Messlinger, K., & Carr, R. W. (2012). Repetitive activity slows axonal conduction velocity and concomitantly increases mechanical activation threshold in single axons of the rat cranial dura. *The Journal of Physiology C*, *590*, 725–735.

[6] Chéreau, R., Saraceno, G. E., Angibaud, J., Cattaert, D., & Nägerl, U. V. (2017). Superresolution imaging reveals activity-dependent plasticity of axon morphology linked to changes in action potential conduction velocity. *Proceedings of the National Academy of Sciences of the United States of America*, *114(6)*, 1401–1406.

[7] Schnitzler, A., & Gross, J. (2005). Normal and pathological oscillatory communication in the brain. *Nature Reviews Neuroscience*, *6*(4), 285–296.

[8] Costa, L. da F., Kaiser, M., & Hilgetag, C. C. (2007). Predicting the connectivity of primate cortical networks from topological and spatial node properties. *BMC Systems Biology*, *1*(1), 16.

[9] Yamazaki Y, Fujiwara H, & Kaneko K et al (2014) Short- and long-term functional plasticity of white matter induced by oligodendrocyte depolarization in the hippocampus. *Glia 62*:1299–1312.

[10] Mckenzie, I. A., Ohayon, D., Li, H., Paes De Faria, J., Emery, B., Tohyama, K., & Richardson, W. D. (n.d.). Motor skill learning requires active central myelination. *Science 346(6207)*, 318-322.

[11] Isaacson, J. S., & Scanziani, M. (2011). How Inhibition Shapes Cortical Activity. *Neuron*, *72*(2), 231–243.

[12] Izhikevich, E. M. (2003). Simple Model of Spiking Neurons. *IEEE Transactions on Neural Networks*, *14*(6).

[13] Song, S., Sjöström, P. J., Reigl, M., Nelson, S., & Chklovskii, D. B. (2005). Highly Nonrandom Features of Synaptic Connectivity in Local Cortical Circuits. *PLoS Biology*, *3*(3), e68.

[14] Tomasi, S., Caminiti, R., & Innocenti, G. M. (2012). Areal Differences in Diameter and Length of Corticofugal Projections. *Cerebral Cortex*, *22*(6), 1463–1472.

[15] Ginzburg I. & Sompolinsky H., (1994) Theory of correlations in stochastic neural networks. *Phys. Rev. E 50*, 3171.

[16] Schnitzler, A., & Gross, J. (2005). Normal and pathological oscillatory communication in the brain. *Nature Reviews Neuroscience*, *6*(4), 285–296.

[17] Chorghay Z., Karadottir R., & Ruthazer E. (2018) White Matter Plasticity Keeps the Brain in Tune: Axons Conduct While Glia Wrap. *Frontiers in Cellular Neuroscience 12*

[18] Polykretis, I., Ivanov, V., and Michmizos, K. P. (2018). A Neural-Astrocytic Network Architecture: Astrocytic calcium waves modulate synchronous neuronal activity. Proceedings of the International Conference on Neuromorphic Systems, 6, pp. 1-8, Nashville, TN, USA.

[19] Polykretis, I., Ivanov, V., and Michmizos, K. P. (2018). The Astrocytic Microdomain as a Generative Mechanism for Local Plasticity. International Conference on Brain Informatics, pp. 1-10, Arlington, TX, USA.